\documentclass[preprint,preprintnumbers,aps,nofootinbib]{revtex4}
\usepackage{amsmath,amssymb}
\usepackage{graphicx}
\usepackage{dcolumn}
\usepackage{bm}

\def \lsim{\mathrel{\vcenter
     {\hbox{$<$}\nointerlineskip\hbox{$\sim$}}}}

\newcommand{\beq}{\begin{equation}}
\newcommand{\eeq}{\end{equation}}
\newcommand{\beqa}{\begin{eqnarray}}
\newcommand{\eeqa}{\end{eqnarray}}
\newcommand{\beqar}{\begin{eqnarray*}}
\newcommand{\eeqar}{\end{eqnarray*}}

\begin{document}

\title{Stealth gluons at hadron colliders}

\author{R.~Barcel\'o}
\email{rbarcelo@ugr.es}
\author{A.~Carmona}
\email{adrian@ugr.es}
\author{M.~Masip}
\email{masip@ugr.es}
\author{J.~Santiago}
\email{jsantiago@ugr.es}
\affiliation{
CAFPE and
Depto.~de F{\'\i}sica Te\'orica y del Cosmos, Universidad de Granada,
18071 Granada, Spain
}

\preprint{UG-FT-287/11}
\preprint{CAFPE-157/11}


\begin{abstract}
We find that a heavy gluon $G$ of mass $800$--$900$ GeV
with small, mostly axial-vector couplings to the 
light quarks and relatively large vector and axial-vector
couplings to the top quark can explain the $t \bar{t}$ forward-backward 
asymmetry observed at the Tevatron with no 
conflict with other top-quark or dijet data. The key ingredient is 
a complete treatment of energy-dependent width effects and  
a new decay mode $G\rightarrow q Q$, 
where $q$ is a standard quark and $Q$ a vector-like quark of mass 
$400$--$600$ GeV.
We show that this new decay channel makes the heavy gluon
\textit{invisible} in the $t\bar{t}$ mass invariant distribution and
discuss its implications at the Tevatron and the LHC.
\end{abstract}


\maketitle

\section{Introduction}
The Fermilab Tevatron is probing quark
interactions up to energies $\sqrt{\hat{s}}\lesssim 800$ GeV.
Its most significant discovery has been the top quark,
with a mass around 172 GeV. This quark could turn out to be
more than {\it just} the building block completing the
third family of the standard model (SM). Since masses are not 
gauge invariant and it is the heaviest fermion, 
one expects that the top-quark sector
holds the key to understand the mechanism of electroweak (EW) 
symmetry
breaking. In particular, new particles (a scalar partner,
a vector-like T quark) or a different (composite) nature  
have been proposed to explain the stability of the EW scale
under top-quark radiative corrections. 

In fact, CDF and D0  
measurements imply an intriguing
deviation with respect to the SM prediction in the
$t \bar{t}$ forward-backward asymmetry $A^{t\bar t}$~\cite{AFBTEV}. 
We will take as reference value  the one recently reported by the CDF
collaboration, 
\beq
A^{t\bar t} \approx \left\{
\begin{array}{l l} 
\displaystyle -0.116 \pm 0.153, \quad
& m_{t\bar t}<450\;{\rm GeV}\,; \\
\phantom{-}0.475\pm 0.114,
& m_{t\bar t}>450\;{\rm GeV}\,,
\end{array} \right. 
\eeq
which refers to the asymmetry measured
in the $t\bar{t}$ center of mass frame.
The SM prediction, $0.040\pm 0.006$ and $0.088\pm 0.013$ for
the low and the high-energy regions, respectively, gives
a three-sigma deviation at large values of 
$m_{t\bar t}$ \cite{AFBTEV}. 
If caused by new physics, this unexpected result would be 
an order-one departure from the standard quark physics  
at 450--800 GeV, and similar
anomalies could be expected in other observables at
the Tevatron and the early LHC. However, the asymmetry 
has not been {\it supported} by current data on 
the total $t\bar t$ cross section or the invariant-mass 
distributions of top-quark pairs and dijets. As a consequence,
possible new particles proposed to explain it are typically 
pushed above 1--2 TeV, 
out of reach both from Tevatron energies and from the
current LHC luminosity. Such high values, in turn, become 
ineffective to produce the large asymmetry or should be apparent
with a slightly increased LHC luminosity, as general effective
Lagrangian studies~\cite{efflag} indicate.

In this letter we show that a heavy gluon with mass
$800$--$900$ GeV can still explain the observed asymmetry with no
conflict with current data. The mechanism that could hide the 
particle responsible for the asymmetry relies on a very large width
caused by new decay channels opening at $\sqrt{\hat s}\lsim 600$
GeV. This requires a careful treatment of the energy-dependent width
of the heavy particle.
The scenario predicts a predominantly right-handed polarization 
for the excess of forward top quarks and at least an extra 400--600 GeV new
quark. Another interesting test of the model is the value of the
$t\bar{t}$ charge asymmetry $A_C$ at the LHC~\cite{AFBLCH}, since
the relatively low mass of 
the gluon resonance implies a change in the sign
at $m_{t\bar t}\approx M_G$. 
Our framework is natural in holographic Higgsless
models~\cite{Barcelo:2011fw}, in which longitudinal $W,Z$  
scattering is unitarized by vector excitations of mass below 1
TeV~\cite{Higgsless}.
In that case our results would imply strongly coupled physics right 
above the EW scale and no Higgs at the LHC.

\section{The model}
The framework is defined by a massive gluon $G$ with large couplings 
to the right-handed top quark, and small-close to
axial ($g^q_L\approx -q^q_R$) couplings to the light 
quarks:
\beqa
g_R^q&=&-(0.2\!\!-\!\!0.3)\,g, \quad 
g_L^q=+(0.2\!\!-\!\!0.3)\,g, \nonumber\\
g_R^t&=&+(4\!\!-\!\!6)\,g, \quad \quad \,\,\,
g_L^t\approx 0\,g.
\label{gencoup}
\eeqa
Couplings in this range are naturally obtained in holographic models 
after imposing consistency with precision EW bounds. 
In those models $g_R^t$ {\it must} be large 
because the top lives in the brane that breaks the EW symmetry,
together with all the massive excitations. 
The opposite sign of $g_R^q$ and $g_L^q$ and the small value
of these couplings are required to reproduce 
the standard coupling with the EW gauge bosons. 
In turn, the sign difference optimizes the appearance of
a FB asymmetry, whereas the size prevents an excess of dijet 
events mediated by the massive gluon. The first feature also implies
that at $\sqrt{\hat s}=m_{t\bar t} \ll M_G$ the contributions 
to ${\rm d}\sigma /{\rm d}m_{t\bar t}$ from 
light quarks of different chirality tend 
to cancel each other \cite{Barcelo:2011fw}.

We take a gluon mass $M_G=800$--$900$ GeV, as required in
Higgsless models and hinted by Tevatron data on $A^{t\bar t}$. 
Finally, the {\it new} ingredient of the set up is 
the presence of a massive quark excitation $Q$ that opens a
new decay channel for the massive gluon or a new gluon-mediated
process at $\sqrt{\hat s}\approx 600$ GeV:
\beq
q\bar q\rightarrow G\rightarrow Q \,\bar q \;,
\label{gt}
\eeq
If $Q$ is an excitation $T$ of the $t$ quark, it should have 
a relatively low mass, $m_T=400$--$500$ GeV. However, if it
decays predominantly into $Wb$ it will produce the same
$WWb\bar b$ signal as $t\bar t$ production and could give
an unobserved effect at $m_{t \bar t}\ge 600$ GeV. The same 
contribution could be obtained from 
an excitation $B$ of the $b$ quark that decays 
significantly into $W t$ (see discussion below). 
Due to the different kinematics, 
a definite statement about the visibility of these final states in
current data requires a detailed analysis that is beyond the scope
of this letter and will be presented elsewhere.
Therefore, we consider the {\it cleanest}
possibility, namely the presence of
one or several heavy quarks $Q$ that are excitations of
the light quarks ($Q=U,D,S,C$) and decay into
$W/Z$ plus jet. The mass of these quarks should be below 
$600$ GeV. To simplify our analysis we will assume a single
particle $Q=U$ and all other quark excitations above the
production threshold in $G$ decays ($m_Q+m_q> M_G$).
We would like to emphasize that the results for other chocies of $Q$
in terms of pure $t\bar{t}$ production are very similar, 
although their collider implications can vary from one case to the
other. 

\section{Role of the $Q$ quark}
Let us start describing the situation in absence of the
the extra $Q$ quark.
It has been shown \cite{Barcelo:2011fw} that the gluon mass 
and the couplings in Eq.~(\ref{gencoup}) are able to produce a large 
FB asymmetry (within 1.5 sigmas of the measured value at
$m_{t\bar t}\ge 450$ GeV) consistently with
bounds from dijet searches. However, there seems to be some 
tension with the data on ${\rm d}\sigma /{\rm d}m_{t\bar t}$ at 
$m_{t\bar t}\ge 600$ GeV, near the gluon mass. This tension is
weak at the 5.3 fb$^{-1}$ Tevatron, but becomes more 
clear at the 0.2 fb$^{-1}$ LHC, where the peak of the 
gluon resonance should be visible. One could hope that 
an increase in the coupling $g_R^t$ will increase the 
gluon width and smear out the peak. However, the fit does not
improve because the total cross section also goes up with
$g_R^t$ and, most notably, the asymmetry $A^{t\bar t}$
seems to decrease.~\footnote{This is the main argument behind recent
  claims in the literature that heavier axigluons are favoured over
  lighter ones~\cite{otheraxigluon,Djouadi:2011aj}.} 
This last effect can be understood because
the gluon width appears in the denominator of the 
$t\bar t$ production amplitude,
and larger values (similar to its mass) 
will suppress the effects of the massive gluon and thus the
predicted asymmetry. 

However, as we increase the gluon width the fixed-width
approximation, standard in current MonteCarlo generators, 
becomes worse. Its energy dependence 
(see for example \cite{Bardin:1988xt,Djouadi:2011aj}) 
can be easily computed from the 
imaginary part of the gluon 2-point function.
We get
\beq
\Gamma_G^{t\bar t}(\hat s)=
 {g^2 \over 24 \pi} {\hat s \over M_G} 
\left(1 - {4 m_t^2\over \hat s }\right)^{1\over 2}
\left[ \left( 1 - {4 m_t^2\over \hat s } \right) g_A^{t\,2}  
+  \left( 1 + {2m_t^2\over \hat s } \right) g_V^{t\,2} \right]
\theta(\hat s-4m_t^2)\;. 
\label{Gtt}
\eeq
In our simulations we have used MADGRAPH/MADEVENT
v4~\cite{Alwall:2007st}, with PYTHIA~\cite{pythia} for 
hadronization/showering effects and PGS4~\cite{PGS4} for detector 
simulation. We have modified the matrix element in the
fortran code generated by MADGRAPH to
include the energy dependence of the width.
This correction tends to increase the effects of
the new physics at low energies. For example, for $g^t_R=6g$ 
the asymmetry at $m_{t\bar t}>450$ goes from 0.17 in a 
simulation with a constant
width to 0.20 using the widht given in (\ref{Gtt}). However, 
we have found that the effect in 
the invariant mass distribution
is small, and the change in slope at $m_{t \bar t}\ge 600$ GeV
would still conflict with the data.

\begin{figure}
\includegraphics[width=.7\linewidth]{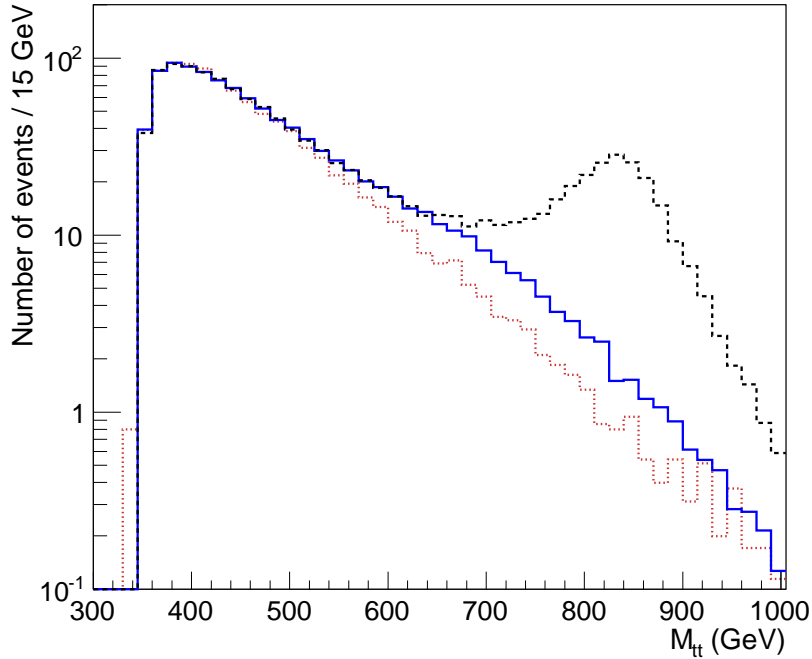}
\caption{\label{fig:TeV} Prediction for the $m_{t\bar{t}}$
  distribution at the Tevatron with a luminosity of $5.3$
  fb$^{-1}$ for the SM (dotted) and for the model
  defined in Eq. (\ref{modelII}) with (solid) and without (dashes)
  the new channel $G\to U \bar u$.}
\end{figure}

The effect of the extra $Q$ quark is then crucial, by opening
the new decay channel $G\to q\bar Q\,,\,\bar q Q$
at $\sqrt{\hat s}\approx 600$ GeV. Below those energies the process
is irrelevant (it does not contribute to the imaginary part of
the propagator), so the FB asymmetry at 450--600 GeV is
unchanged. At $m_{t\bar t}\ge 600$ GeV, in contrast, this decay
channel will {\it dissolve} the peak without increasing the 
number of $t\bar t$ pairs produced. Its energy-dependent 
partial width is
\beqa
\Gamma_G^{q\bar Q,Q\bar q}(\hat s)&=&
 {g^2 \over 12 \pi} {\hat s \over M_G} 
\left(1 - {(m_q+m_Q)^2\over \hat s} \right)^{1\over 2} 
\left(1 - {(m_q-m_Q)^2\over \hat s} \right)^{1\over 2}\times
\nonumber \\ 
&&
\left[ \left( 1 - {m_q^2+m_Q^2+6 m_q m_Q\over 2 \hat s } -
{(m_Q^2 - m_q^2)^2\over 2 \hat s^2 } \right) \right. g_A^{Q\,2} +
\nonumber \\
&&
\left. \left( 1 - {m_q^2+m_Q^2-6 m_q m_Q\over 2 \hat s } -
{(m_Q^2 - m_q^2)^2\over 2 \hat s^2 } \right)g_V^{Q\,2}  \right] 
\theta(\hat s-(m_q+m_Q)^2)\;.
\label{GtT}
\eeqa
We take as a benchmark model the following values of the parameters
\begin{eqnarray}
M_G&=&850\mbox{ GeV}, \quad M_U=500\mbox{ GeV}, \nonumber \\
g_L^q&=&-g_R^q=-g_R^b=0.3 g, \quad \nonumber \\
g_L^b&=&g_L^t=0, \quad g_R^t=4 g, \nonumber \\
g_R^{uU}&=& 7.6 g, \quad \mbox{ all other couplings $=$ 0}
\label{modelII}
\end{eqnarray}
We have chosen the coupling of the heavy gluon to $uU$ in such a way
that the total width at the gluon mass is $\Gamma_G=0.7\,M_G$.
We plot in Fig.~\ref{fig:TeV} the event distribution for this model
with and without the $uU$ channel included (solid and dashes,
respectively) together with the SM prediction (dotted line). 
It is clear from the
figure that without the new channel 
the peak is clearly visible. Once it
is included the large width makes the gluon completely invisible.
Including the SM contribution, the FB asymmetry in the large
$m_{t\bar t}$ region goes from $A^{t\bar t}=0.30$ with
no extra $U$ quark to $A^{t\bar t}=0.33$ in this 
model, just $1.2 \, \sigma$ away from the measured
value. 

\begin{figure}[t]
\includegraphics[width=.7\linewidth]{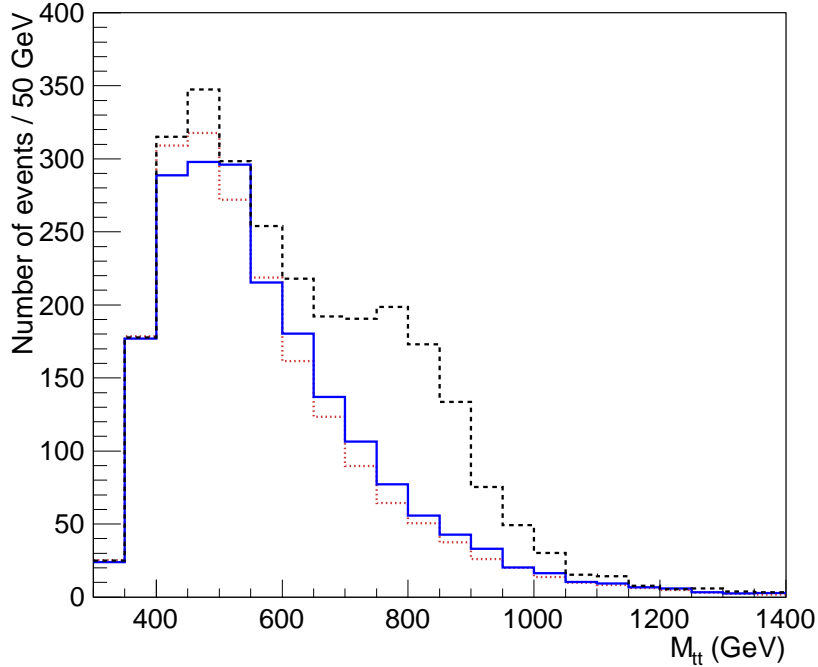}
\caption{\label{fig:LHC} Prediction for the $m_{t\bar{t}}$
  distribution at the LHC with a luminosity of $1$
  fb$^{-1}$ for the SM (dotted) and for the model
  defined in Eq. (\ref{modelII}) with (solid) and without (dashes)
  the new channel $G\to U \bar u$.}
\end{figure}

It is then clear that our \textit{stealth}-gluon model can reproduce 
the Tevatron
data on the forward-backward asymmetry and $m_{t\bar{t}}$ distribution.
Analogous observables could also test the model at the 
LHC~\cite{LHCttbar}. We show in Fig.~\ref{fig:LHC} the expected
$m_{t\bar{t}}$ distribution there for 1 fb$^{-1}$ of
integrated luminosity. We have followed the analysis in the first reference
in~\cite{LHCttbar} and taken the 4 jet, 2 $b$ tags in the muon channel as an
example. It is clear from the figure that, given the uncertainties in
the $t\bar{t}$ normalization, this gluon is also invisible at the
LHC. 

\section{Other implications at hadron colliders}
As we have already mentioned, the small couplings (0.2--0.3)$g$
of the massive gluon to the light quarks implies an acceptable 
contribution to dijet data. However, a 400--600 GeV
new quark $Q$ can be searched for both at the Tevatron
and the LHC. Standard searches are based on either (QCD) pair
production or single production through electroweak
interactions. Current bounds are in the $m_Q \lesssim
385$ GeV region, depending on its preferred decay channel~\cite{Qexperiment}. 
Single production through $G$
in the $s$--channel together with a SM quark is therefore
a novel mechanism. 
A detailed analysis of all possible decay channels and how competitive
this single production can be with more standard searches is beyond
the scope of this work and will be deferred to a forthcoming
publication. 
(See~\cite{Burdman:2010gr} for a related discussion in the context of
heavy gluons with fourth generation models.)
It is important to emphasize, however, that some of these channels might
produce non-standard signals that could be easy to miss if not
explicitely searched for. 
As an example, if we take the case of a $B$
with mostly charged current decays the final state would be exactly
the same as that of $t\bar{t}$. The signal could then be missed
because simple searches that do not use $t$ reconstruction peak at
$\sim 500$ GeV, which is precisely where the peak in our model would
show up, and more sophisticated analyses that include reconstruction
could miss our signal because the two $W$ come from the same leg.
The case with lighter quarks, for example 
\beq
q \bar q \to G\to U \bar u \to W d \bar u
\eeq
could be searched as $W$ plus 2 jets.~\footnote{If the coupling $VQq$
with $V=Z,W$ is large both the Tevatron and the LHC have good chances
of discovering the new quark in single production~\cite{Usingle}.}

Another consequence of this scenario is the polarization of
the top quarks produced at large 
values of $m_{t\bar t}$. In the model that we are considering
the left-handed top quark does not couple to $G$. As a consequence,
the forward (backward) excess (deficit) is defined 
basically by right-handed top quarks.  
The polarization can be measured in the 
subsequent decay $t\to b\,l^+\nu_l$ from the distribution of the final
lepton (see for example~\cite{Agashe:2006hk}).

Finally,  the charge asymmetry in $t\bar t$ production at large 
rapidities 
can be measured at the LHC in a wide range of $m_{t\bar
  t}$~\cite{AFBLCH}. Our
scenario predicts that the asymmetry changes sign at 
$m_{t\bar t}=800$--900 GeV. Notice that if the Tevatron could
measure $A^{t\bar t}$ above the gluon mass it should also find  
a slight forward deficit. 
This very small and even negative
charge asymmetry at large values of $m_{t\bar{t}}$ can be an
important test discriminating our model from other explanations of the
Tevatron $A^{t\bar{t}}$ (see~\cite{AguilarSaavedra:2011hz} for a
discussion in terms of effective operators).

\section{Summary and discussion}
The large FB asymmetry observed at the Tevatron, if confirmed, is  
a sign of new physics at 450--800 GeV. This physics can compete
with the color interaction, which suggests a strong coupling and
a relatively light mass for the mediator. On the other hand, the
absence of a peak in the $t\bar t$ invariant mass distribution
or of an excess of dijets at hadron colliders seem to imply
that the new physics should be
weaker and heavier, making difficult the definition of a working
model. We have shown that all observations can be simultaneously
satisfied if the mediator is a 800--900 GeV gluon with 
an additional decay mode into 
a SM quark $q$ plus an extra $Q$ quark of 400-600 GeV. 
The new channel does
not change the physics below $m_q+m_Q$, preserving the FB asymmetry,
but it suppresses the effects of the gluon
at higher $m_{t\bar t}$.
The signal from the $Q$ quark is model dependent. In holographic
models it could be one or several excitations of the light quarks
or it may be
a resonance of the right handed top-quark that decay mostly
into $Zt$ (or even $Ht$ in models
with a light Higgs). Some of the different possibilities for $Q$
give
rise to novel phenomenology that could be missed by standard LHC
searches unless explictly tailored analyses are used.

Another interesting feature of our setup is that the 
top-quark Tevatron excess in the forward direction is mostly
composed of right-handed top quarks, which could be tested
studying the angular distribution of the positron resulting
from their decay. At the LHC, the charge asymmetry
in $t\bar t$ production at large rapidities
should change sign at $m_{t\bar t}\approx M_G$. 

We think that the best fit for the unexpected data on 
the FB asymmetry at the Tevatron
is strongly coupled physics below 1 TeV.
Our scenario is naturally realized in holographic Higgsless
models. Thus, a clear consequence of such scenario at the
LHC would be the absence of the light Higgs preferred by
the SM or its SUSY extensions and no new physics up to
energies around $m_q+m_Q\approx 600$ GeV.

\begin{acknowledgments}
We would like to thank G. Azuelos, F. Boudjema, G. Brooijmans,
M. Felcini, U. Haisch, J.I.~Illana, 
G. Moreau and S. Westhoff for useful
discussions. 
This work has been partially supported by
MICINN of Spain (FPA2006-05294, FPA2010-16802, FPA2010-17915,
Consolider-Ingenio 
{\bf Multidark} CSD2009-00064, and Ram\'on y Cajal Program)  
and by Junta de Andaluc\'{\i}a
(FQM 101, FQM 3048 and FQM 6552).
\end{acknowledgments}

\end{document}